\documentclass{raa}
\usepackage[ngerman,english]{babel}
\newif\ifAMStwofonts
\AMStwofontstrue 
\def\degr{\hbox{$^\circ$}}

\def\arcsec{\hbox{$^{\prime\prime}$}}

\def\aobs{{\alpha_{\rm _{obs}}}}
\def\baobs{{{\bar\alpha}_{\rm _{obs}}}}

\def\gsim{~\rlap{$>$}{\lower 1.0ex\hbox{$\sim$}}}

\def\simpropto{\lower.2ex\hbox{$\; \buildrel \propto \over \sim \;$}}
\def\ltsim{\lower.5ex\hbox{$\; \buildrel < \over \sim \;$}}
\def\gtsim{\lower.5ex\hbox{$\; \buildrel > \over \sim \;$}}
\def\ltsim{\lower.5ex\hbox{$\; \buildrel < \over \sim \;$}}
\def\gtsim{\lower.5ex\hbox{$\; \buildrel > \over \sim \;$}}

\def\no{\noindent}

\def\ln{{\rm ln}}

\def\pmb#1{\setbox0=\hbox{#1}%
\kern-.025em\copy0\kern-\wd0
\kern.05em\copy0\kern-\wd0
\kern-.025em\raise.0433em\box0}

\def\vs{\pmb{s}}

\def\simlt{\lower.5ex\hbox{$\; \buildrel < \over \sim \;$}}
\def\simgt{\lower.5ex\hbox{$\; \buildrel > \over \sim \;$}}

\newcommand{\beq}{\begin{equation}}

\newcommand{\eeq}{\end{equation}}
\def\beqa{\begin{eqnarray}}
\def\eeqa{\end{eqnarray}}
\def\fixit#1{}

\def\aaps{A\&AS}
\def\apjs{ApJ. S}

\def\apjl{ApJL}

\def\aap{A\&A}

\def\apjl{ApJL}

\usepackage{graphicx,times} 
\usepackage[colorlinks = true,linkcolor = blue,urlcolor= blue,citecolor= cyan,anchorcolor = blue]{hyperref}
\usepackage{natbib}
\usepackage{breakurl}
\usepackage[latin9]{inputenc}
\usepackage[T1]{fontenc}
\usepackage{amssymb}
\usepackage{amsmath}
\usepackage{multirow}
\usepackage{mathrsfs}
\usepackage{bbding}
\usepackage{soul,xcolor}
\usepackage{float}
\usepackage{color}
\setstcolor{red}
\bibpunct{(}{)}{;}{a}{}{,}
\newcommand\wx{1.0}
\begin{document}
\title{Radio spectral index from NVSS and TGSS}
\volnopage{Vol.0 (20xx) No.0, 000--000}      
\setcounter{page}{1}   
\author{Prabhakar Tiwari}
\inst{1}
\institute{National Astronomical Observatories, Chinese Academy of Sciences \\
 Beijing, 100012, China\\ 
ptiwari@nao.cas.cn}
\abstract{
I extract the radio spectral index, $\alpha$, from  541,195 common sources observed in the 150 MHz TIFR GMRT Sky 
Survey (TGSS) and the 1.4 GHz NRAO VLA Sky Survey (NVSS). This large   common source catalogue covers 
about $80\%$ of the sky. The flux density limits in these surveys are such that the observed galaxies are  presumably hosts 
of active galactic nuclei (AGNs). I confirm the steepening of $\alpha$ with increasing flux density for this 
large sample and provide a parametric fit between $\alpha$ and flux density. Next, I divide the data into a 
low flux (LF) and a high flux (HF) density sample of roughly equal number of galaxies. The LF sample contains all 
galaxies below 100 mJy TGSS and 20 mJy NVSS flux density and the HF sample is all galaxies 
above  100 mJy TGSS and 20 mJy NVSS. I observe an increase 
in $\alpha$ with source size (TGSS measured),  saturating  for large sizes to $0.89\pm0.22$ and $0.76\pm 0.21$ for 
the LF and  HF sources, respectively. I discuss the observed results and possible physical mechanisms 
to explain observed $\alpha$ dependence with source size for LF and HF samples. }
\maketitle

\section{Introduction}
\label{sec:int}
The radio emission from a distant galaxy consists of thermal and  synchrotron contributions. The  thermal emission 
is bremsstrahlung radiation from  H{\scriptsize II} regions while the non-thermal synchrotron emission is generated by
supernova remnants (SNRs), diffuse cosmic ray electrons (CREs)
spread over the disk and halo \citep{Biermann:1976,Condon:1992}, relativistic plasma in 
jets and jet fed lobes \citep{Young:1976,Blandford:1979}. The bright radio galaxies, 
relevant to flux density limits in this work, are hosts of  AGNs and their radio emission is predominantly 
synchrotron radiation from jets of relativistic plasma and jet fed 
lobes\footnote{ and the inverse-Compton scattering of the radio synchrotron emission within the lobes}. 
These emission components are characterized by  different spectral indices and, therefore,  the total 
spectral index depends on their relative contributions. The thermal  emission is almost flat
($\propto \nu^{-\alpha}$) with  spectral index $\alpha_{\rm_{th}}\approx 0.1$\citep{Condon:1992}, whereas, the 
radio jets exhibit a  spectral index $\alpha_{\rm _{jet}} =0.5-0.7$ \citep{Bridle:1984, Laing:2013} and the radio emission from 
lobes can be flat to very steep depending on energy injection and  losses \citep{Kellermann:1966}.

I compile a common source catalogue from  TIFR GMRT Sky Survey first alternative data release  
(TGSS ADR1) \citep{Intema:2016tgss}  and NRAO VLA Sky Survey (NVSS)\citep{Condon:1998} catalogues, and use it  
to explore radio spectral index, size, and flux density dependence. I identify 
a  total of 541,195  sources  which are common to  both TGSS ADR1 and NVSS. The large sample allows us to 
separate spectral index values in to differential source size bins and I obtain statistically significant dependence of spectral 
index on flux density and size. Assuming that the NVSS source population peaks between 
redshift $z\approx0.5-1$ \citep{Wilman:2008, Adi:2015nb,Tiwari:2016adi}, all results in this work statistically 
represents the radio source population at redshift  $z\approx 0.8$. 

The outline of the paper is as following. I discuss the NVSS and TGSS ADR1 data and describe the cross-matching 
procedure in Section \ref{sc:data}. In  Section \ref{sc:theory} I discuss the basic mechanisms of radio 
emission and their spectral index. I describe the observed spectral index in Section \ref{sc:alp_obs}. 
In Section \ref{sc:res} I present all results. I conclude with discussion in Section \ref{sc:summary}.

\section{Data}
\label{sc:data}
\subsection{TGSS ADR1}
\label{ssc:tgss}
The TGSS ADR1 catalogue is based on the Giant Metrewave Radio Telescope (GMRT), performing all-sky radio continuum survey 
at 150 MHz \citep{Swarup:1991}.  The catalogue is surveyed  between 2010 and 2012 and  has been published recently  
by \cite{Intema:2016tgss}. The catalogue  covers $~90\%$ of the full sky at 150 MHz with median RMS brightness 
fluctuations 3.5 mJy/beam with approximate resolution 25\arcsec x 25\arcsec north of 19$\degr$ DEC and 
25\arcsec x 25\arcsec/cos(DEC-19$\degr$) south of 19$\degr$. The best resolution observed with GMRT at 150 MHz is 20\arcsec \citep{GMRT} and between declination -20$\degr$ to +60$\degr$ TGSS ADR1 sources are best resolved with resolution close to 20\arcsec (see figure $6$ in \cite{Intema:2016tgss}).
The catalogue contains a total 623,604 sources (489,570 sources above flux density 
\footnote{``flux density" refers to total flux density throughout the paper.} 50 mJy) observed 
at $6\sigma$  peak-to-noise threshold. The accuracy of the source centroid position  is better than 2\arcsec (RMS). 
The catalogue is almost complete above 100 mJy \citep{Intema:2016tgss} and $\sim70$\% complete above 50 mJy at 150 MHz.
\subsection{NVSS}
\label{ssc:nvss}
The  NVSS catalogue contains $\sim$1.8 million sources with flux densities $S_{\rm 1.4GHz}>2.5$ mJy at 1.4 GHz \citep{Condon:1998}. 
The full width at half maximum angular resolution is 45\arcsec and nearly all observations are performed at uniform sensitivity. 
The catalogue covers  about 82$\%$ of the sky and has a 100\% overlap with TGSS ADR1. The RMS uncertainty in angular position 
is up to  7\arcsec, substantially larger than TGSS. The catalogue is complete above 3.5 mJy at 1.4 GHz.

\begin{figure}
\includegraphics[width=\wx\textwidth, angle=0]{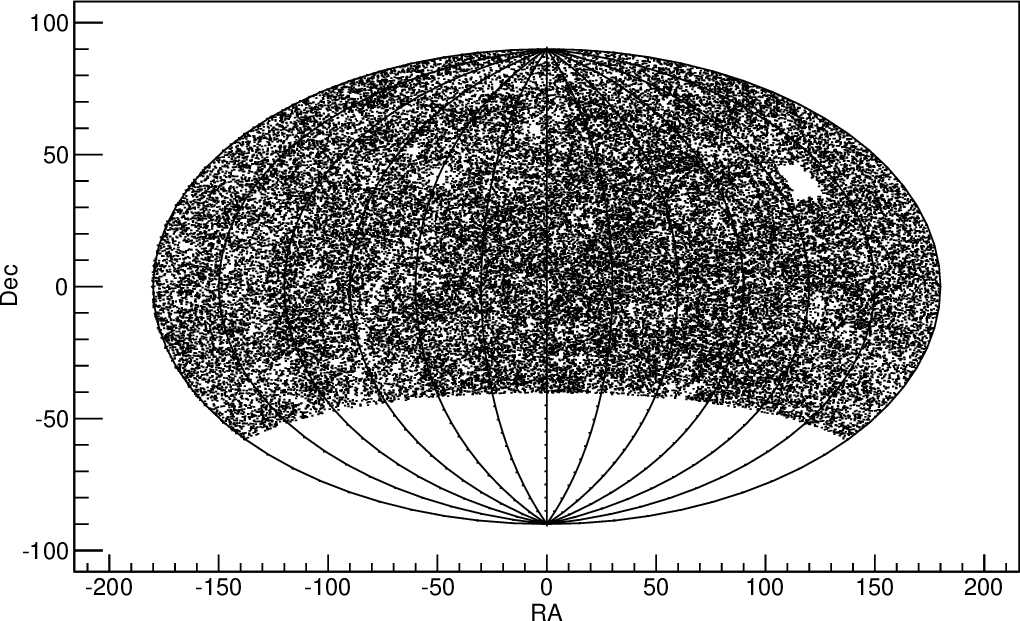}
\caption{The angular distribution of NVSS-TGSS common sources. Only  10\% randomly selected sources are plotted.}
 \label{fig:NVSS_TGSS}
\end{figure}

\subsection{TGSS-NVSS common source catalogue}
\label{ssc:tgss_nvss}

Given the TGSS angular position of each of the 623,604 TGSS ADR1 sources, I identify NVSS sources within an angular  distance 
less than 30\arcsec. In total I find 553,301 TGSS ADR1 sources having at least one NVSS source within 30\arcsec. 
Given the positional uncertainties of NVSS (Section \ref{ssc:nvss}) and TGSS ADR1 (Section \ref{ssc:tgss}) and with 30\arcsec criteria 
we have a less than 4 sigma chance to miss a source. Therefore due to positional uncertainty, we expect only to miss a few 
tens of sources while identifying NVSS match for 0.6 million TGSS ADR1 sources. Furthermore, given the NVSS source number 
density, Section \ref{ssc:nvss}, there is a finite probability, $\approx 0.013$, of having a NVSS source in 30\arcsec radius 
and we expect to misidentify around 8000 sources due to this probability. Due to all these and other possible 
observational systematics and uncertainties \citep{Condon:1998,Intema:2016tgss} I find 12,106 sources, 
out of 553,301 cross-matched, these are either showing more than one NVSS association with search criteria of 30\arcsec 
or more than one TGSS sources showing association with same NVSS object. I 
drop these 12,106 sources to avoid confusion and artificial scatter in the spectral index. I end up with a total 541,195 
sources, identified in both TGSS ADR1 and NVSS. The common source counts for different TGSS flux density cuts are listed in 
Table 1.
The spatial distribution of these common sources is shown in Fig. \ref{fig:NVSS_TGSS}. 
\begin{table} 
\label{tb:TGSS-NVSS}
\caption{TGSS-NVSS common sources with different TGSS flux density cuts.}
\begin{tabular}{@{}cc}
\hline
${\rm S_{TGSS}}$  &  number of sources \\
(mJy)  & \\
\hline
no cut & 541,195 \\
$>50$  & 424,639\\
$>100$ & 266,885 \\
$>150$ & 190,830 \\
$>200$ & 147,171 \\
\hline 
\end{tabular} 
\end{table}

\section{Radio emissions in AGNs}
\label{sc:theory}
\subsection{Thermal radio emission from H{\scriptsize II} regions}
\label{ssc:th}
The flux density of radio thermal emission dependence on frequency is expressed as \citep{Condon:1992}, 
\begin{equation}
\label{eq:alpha_th}
S_{T} \propto \nu^{-0.1} \; . 
\end{equation}
At high frequencies ($\sim$ GHz) where the electron deflection opacity is small, the 
ratio of the non-thermal emission approximate thermal radio emission $S_T$ with respect to non-thermal radio 
emission $S_{\rm NT}$ is  \citep{Condon:1990},
\begin{equation}
\label{eq:St_frac}
 S_{\rm NT} \sim 10 \left( \frac{\nu}{\rm GHz} \right)^{0.1-\alpha} S_T,
\end{equation}
where $\alpha \sim 0.8$ is the approximate  non-thermal spectral index 
\citep{Intema:2011,Williams:2013,Williams:2016,Hardcastle:2016,Mahony:2016b}. This implies that the non-thermal 
emission is dominant with   a  thermal contribution of $\ltsim10$\% at 1.4 GHz and  $\ltsim3$\%  at 150 MHz.

\subsection{Non-thermal radio emission}
\label{ssc:NT}
 The radio galaxies are broadly categorized in \cite{Fanaroff:1974} type I (FR I) and II (FR II), and the large-scale 
structure of a radio galaxy is roughly consist of three components: jets, lobes and hotspots 
\citep{Harwood:2013,Harwood:2015}. In FR I galaxies, the jets are relatively less collimated and produce radio emission 
trough deceleration \citep{Laing:2013,Laing:2014}, whereas the FR II jets are often non-emitting and the radio emission
largely produced in lobes and hotspots \citep{Harwood:2013,Harwood:2015}.

\subsubsection{Radio emission from jets}
\label{ssc:jets}
The relativistic jet transports the material from central AGN to extremities of the source and terminates as 
shock forming a hotspot i.e. a compact region of synchrotron emission. At the point of acceleration, the electrons 
number distribution is expressed as   $N(E)\propto E^{-\gamma}$, where $\gamma \approx 2$ for FR I 
\citep{Blandford:1979} and $\gamma > 2$ for FR II \citep{Harwood:2016,Harwood:2017b}. 
In FR I galaxies, for which the emission form jets is significant, the mean spectral 
index of radio emission from jets decreases with distance from the nucleus and $\alpha_{\rm _{jet}}\approx0.5-0.7$ 
\citep{Laing:2013}.

\subsubsection{Radio emission from Jet fed lobes }
\label{ssc:lobes}
The radio emission from extended radio-loud galaxies is dominant by radio lobes which  are fed by collimated narrow jets. 
The radio emission from these lobes depends on  electron injection and energy losses. 
For a continuous supply of electrons, 
the number distribution obtained by considering losses to  synchrotron is $N(E)\propto E^{-1-\gamma}$ \citep{Kellermann:1966}, corresponding to  
$\alpha =\gamma/2$. If the energy injection is intermittent, then  $\alpha$ depends on the frequency $\nu$ and 
the time interval between the  bursts. In this case, the first-order Fermi 
acceleration assuming strong shocks \citep{Bell:1978,Blandford:1978,Kirk:2000,Lemoine:2003} gives relatively flat spectrum
($\alpha_{\rm _{lobes}}\ge0.5$) at low frequencies and very steep ($\alpha_{\rm _{lobes}}=1.33$) 
at higher frequencies \citep{Kellermann:1966}.

\subsubsection{Radio emission from core}
\label{ssc:core} 
The spectral characteristics of ``compact" (i.e., unresolved at $\sim 1\arcsec$ resolution) radio sources is very different from 
``extended" sources. The radio emission from the core is significant for compact radio sources \citep{Antonucci:1993,Urry:1995} 
and assuming the power-law distribution of electron energies, the radio spectra of compact sources are usually flat with
$\alpha_{\rm _{core}} \leq 0.5$ \citep{Peterson:1997}. In particular for compact blazars the 
radio emission from the core dominates and the radio spectral index  is  $\approx 0.0 $ \citep{Fan:2010}.

\section{Observed spectral index}
\label{sc:alp_obs}
I derive  the spectral index, $\alpha$,  from the measured radio emission  at frequencies $\rm \nu_{_{TGSS}}=$150 MHz  
and $\rm \nu_{_{NVSS}}=$1.4 GHz  for objects in the TGSS-NVSS common source catalogue. For each source I compute the 
observed spectral index as,
\begin{equation}
\label{eq:aobs}
\aobs = \frac{\ln \left({\rm S_{TGSS}}/{\rm S_{NVSS}}\right) }{\ln \left({\rm \nu_{_{NVSS}}}/{\rm \nu_{_{TGSS}}}\right)}\; ,  
\end{equation}
where  ${\rm S_{TGSS}}$ and ${\rm S_{TGSS}}$ are, respectively,  the flux densities measured by the TGSS and NVSS. 
As we have seen in Section \ref{ssc:th}, thermal emission is negligible and 
this $\aobs$ is mainly fixed by  the synchrotron component. 

I divide the catalogue of common source in to two samples. The first low flux (LF) density sample contains  sources 
with TGSS and NVSS flux densities  below 100 mJy and 20 mJy, respectively\footnote{Note that ${\rm S_{TGSS}}=$ 100 mJy 
corresponds to ${\rm S_{NVSS}}\approx $ 20 mJy, assuming spectral index $\approx 0.73$. }. The second high flux (HF) density sample includes sources above  
and equal ${\rm S_{TGSS}} =100$ mJy and ${\rm S_{NVSS}}=20$ mJy. These flux density cuts are imposed so that both 
samples LF and HF contain roughly 50\% of sources from common source catalogue. We notice from 
\cite{Wilman:2008} (see figure $4$ therein) simulation that the LF sample is expected to 
consists of  mostly FR I  galaxies whereas HF sample supposed to contain almost equal number of FR I and FR II galaxies. 
Furthermore, note that the TGSS ADR1 catalogue is approximately complete above 100 mJy and so the HF sample represents 
the full radio source population \citep{Intema:2016tgss}. The LF sample is increasingly incomplete for lower flux densities. Nonetheless, LF 
sample represents the sources with low flux density.

The non-thermal radio emission  consists  mainly of synchrotron emission from jets and lobes. The spectral 
index from such systems  is not  accurately known. For jets the spectral index $\alpha_{\rm _{jet}}$ ranges 
from 0.5 to 0.7 \citep{Laing:2013} and for lobes the nor-thermal emission may be anywhere between 0.5 to 1.33 
\citep{Bell:1978,Blandford:1978,Kirk:2000,Lemoine:2003,Kellermann:1966}. In general we can not isolate the 
radio emission from jets and lobes, however, the relative contribution from 
lobes dominates in large sources. For ultra large sources and for continuous energy injection into the lobes,  
$\alpha_{\rm _{lobes}}$ saturates to the value $\gamma/2$, where $\gamma$ refers to the electrons in the lobes.

\section{Results}
\label{sc:res}
\subsection{Radio spectral index dependence on flux density}
\label{sc:res:Sp_Flux}
\begin{figure}
\includegraphics[width=\wx\textwidth]{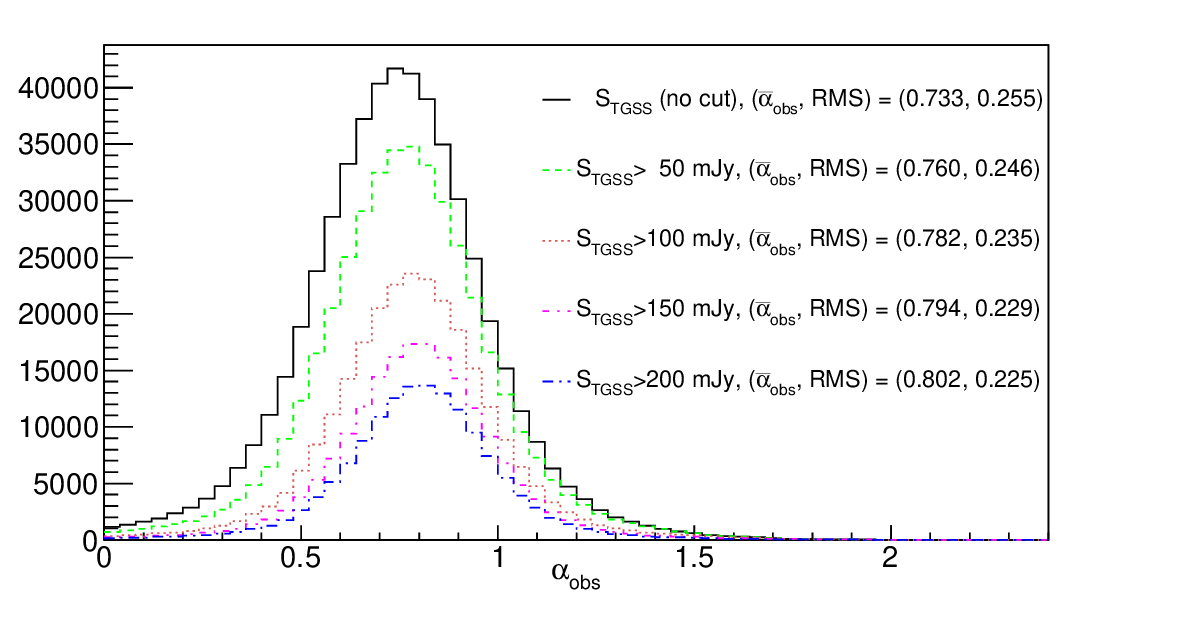}
\caption{The NVSS-TGSS extracted radio spectral index with various TGSS flux density cuts.  
The RMS is the root mean square from $\aobs$ distribution. The TGSS ADR1 is complete above $S_{\rm TGSS}>100$mJy. }
 \label{fig:sp_NT}
\end{figure}

\begin{figure}
\includegraphics[width=\wx\textwidth]{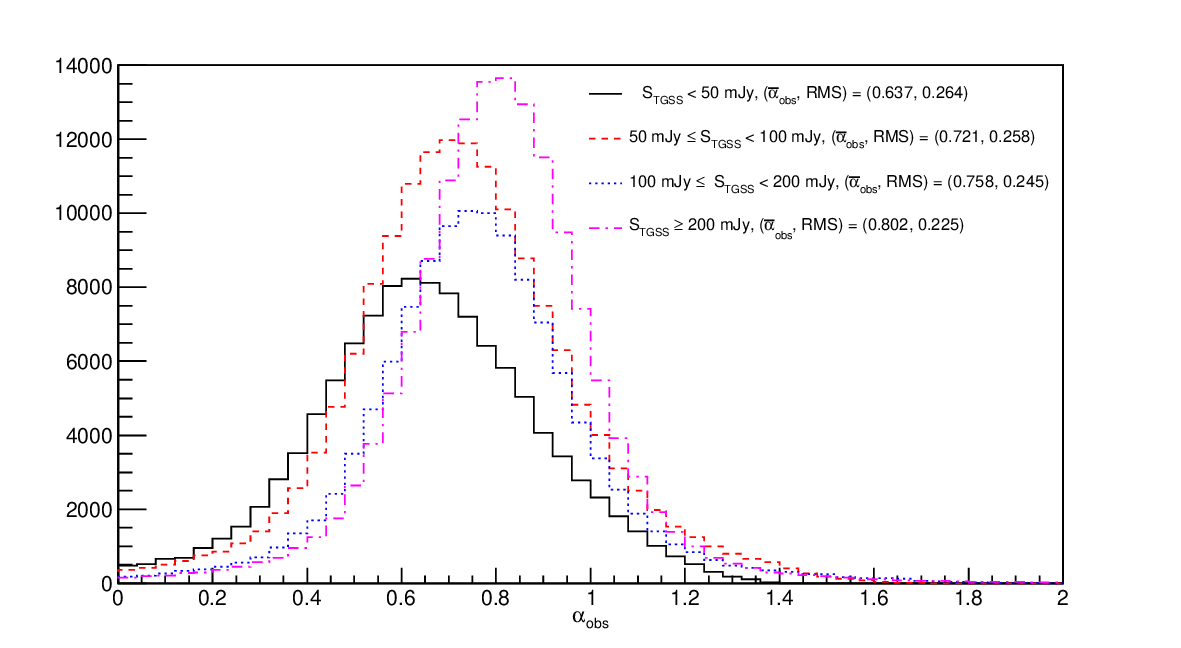}
\caption{The NVSS-TGSS extracted radio spectral index in various TGSS flux density bins.}
 \label{fig:sp_NT2}
\end{figure}

\begin{figure}
	\includegraphics[width=\wx\textwidth]{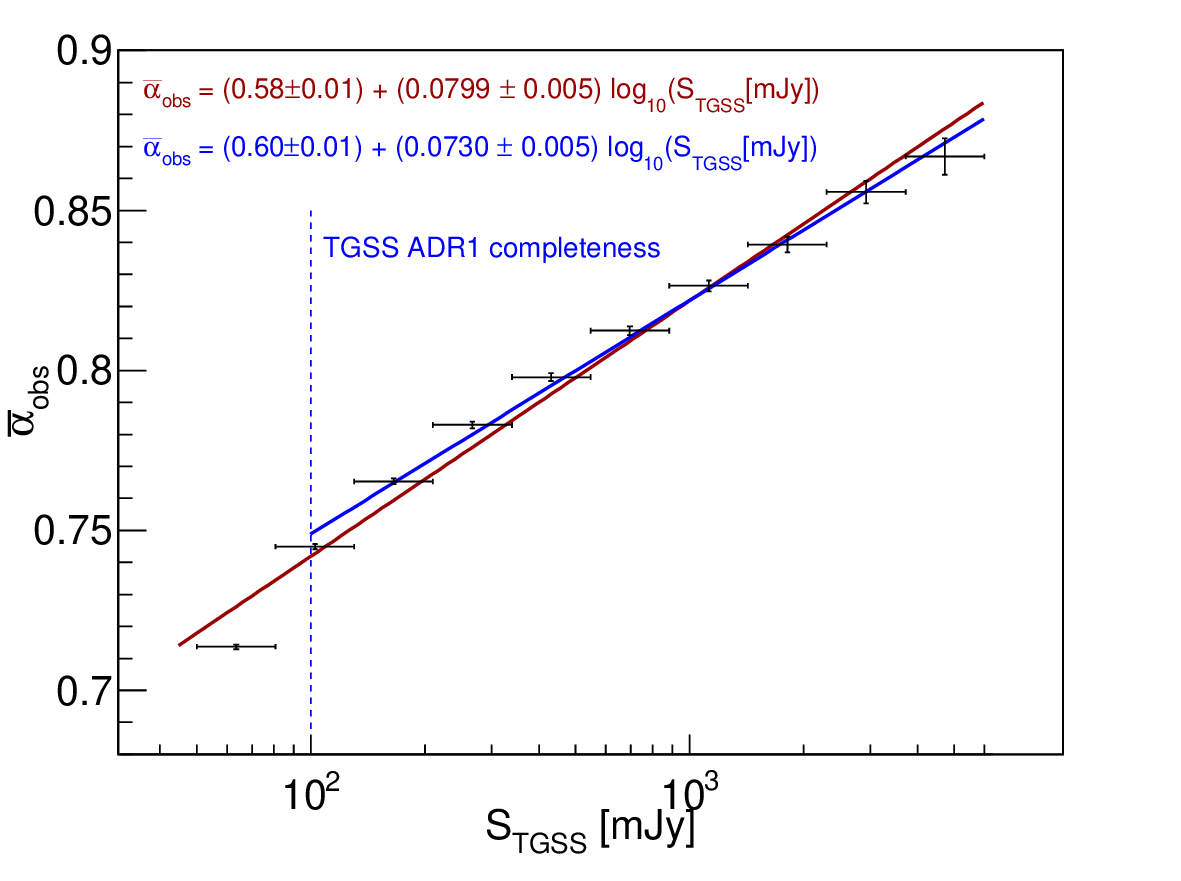}
\caption{The mean radio spectral index, $\baobs$,  as a function of the source TGSS flux density. The vertical error-bars are 
the $1\sigma$ error in $\baobs$ in each flux density bin. The horizontal error-bars represent the width of each bin.  The 
red and blue line represent the fits for full TGSS-NVSS common catalogue and for the common catalogue with TGSS ADR1 flux density 
completeness, respectively. Fit goodness: $\chi^2/ {\rm NDF}$ is 0.9649/7 (blue)  and 5.086/8 (red). 
Corresponding p-values are 0.9954 (blue) and 0.7484 (red). }
\label{fig:sp_S}
\end{figure}

\begin{figure}
\includegraphics[width=\wx\textwidth]{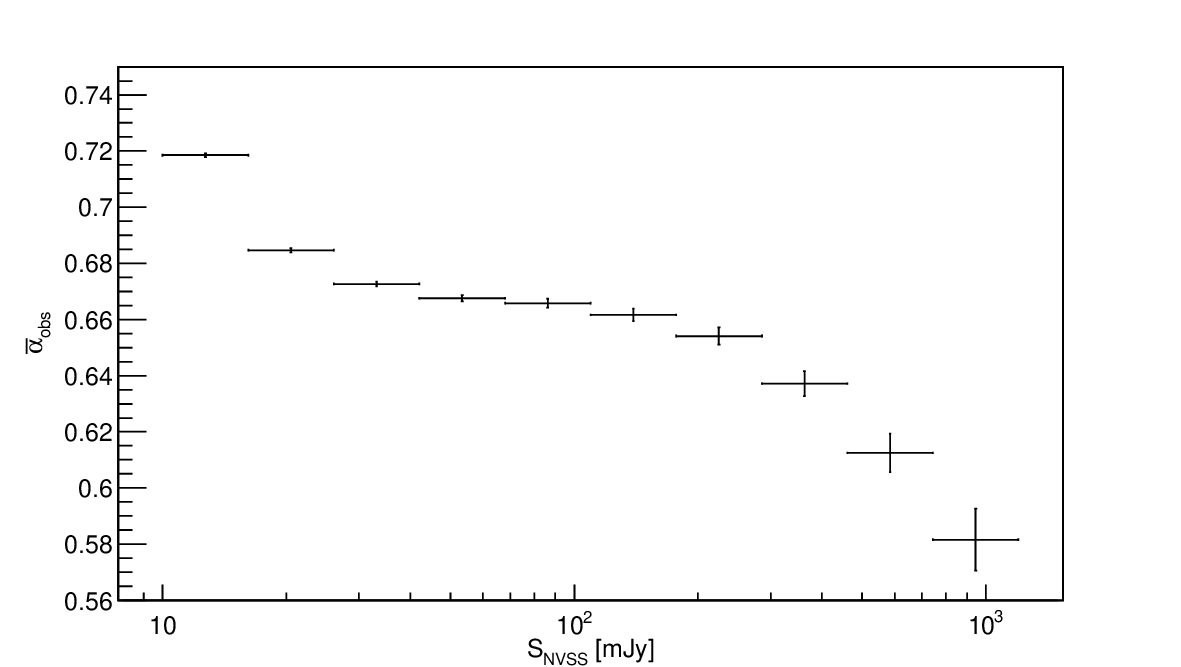}
	\caption{The mean radio spectral index, $\baobs$,  as a function of the source NVSS flux density. The error bars 
	are as in Fig. \ref{fig:sp_S}.}
\label{fig:sp_S2}
\end{figure}

Fig. \ref{fig:sp_NT} and \ref{fig:sp_NT2}  show the distribution of  $\aobs$  for sources with different flux density 
cuts and bins, respectively. Although the RMS spread is large, we see a clear 
trend of steepening of the spectral index  with increasing flux density. The error in 
spectral index measurement due to flux density uncertainty is \citep{Mahony:2016b},
\begin{equation}
\label{eq:err_alpha}
\delta \alpha = \frac{1}{\ln \frac{\nu_1}{\nu_2}} \sqrt{\left(\frac{\delta S_1}{ S_1}\right)^2+ \left(\frac{\delta S_2}{ S_2}\right)^2}
\end{equation}
where $\nu_{1,2}$ and $S_{1,2}$ referes to NVSS and TGSS frequencies and flux densities respectively. 
The flux density accuracy in TGSS ADR1 and NVSS is $\sim10$\% \citep{Intema:2016tgss} and $\sim5$\% \citep{Condon:1998}, respectively. 
Using equation \ref{eq:err_alpha}, these flux density uncertainties give $\delta \alpha \sim0.05$.
However, the RMS spread observed in Fig. \ref{fig:sp_NT} is about $0.24$ which is much larger than that 
expected from flux density measurement uncertainty. Thus the RMS spread observed is largely  the intrinsic 
distribution of source spectral indices. This is expected  since the individual sources have very different 
radio emission properties, i.e. it may be FR I, FR II type and its radio flux may be largely
from jets or lobes. To explore this further, I sort the  sources 
in logarithmic bins of ${\rm S_{TGSS}}$ and  plot the mean spectral index $\baobs$ in each bin as a function of the 
flux density  in Fig. \ref{fig:sp_S}. The observed dependence is fitted  by  linear parametric form as indicated in 
the figure. This confirms the earlier hints of spectral index dependence on flux 
density \citep{Gopal-Krishna:1982,Steppe:1984,Windhorst:1993,Prandoni:2006,Ibar:2009,Ishwara:2010,Randall:2012}.  
However this trend of increasing spectral index is exclusinve to flux density bins of 
${\rm S_{TGSS}}$ i.e. to low frequency bins. The trend is almost inverted if we sort the  sources
in logarithmic bins of ${\rm S_{NVSS}}$ and  plot the mean spectral index (Fig. \ref{fig:sp_S2}). This 
is expected and evident from equation \ref{eq:aobs}, $\frac{d\aobs}{d \ln {{\rm S_{TGSS}}}}$ and 
$\frac{d\aobs}{d \ln {{\rm S_{NVSS}}}}$ are positve and negative, respectivelly.

The TGSS ADR1 data has high noise in the Galactic plane \citep{Intema:2016tgss}; furthermore, there are compact 
bright source present in the Galactic plane \citep{White:2005} with significant thermal emission. Therefore, I
mask the Galactic plane to check the robustness of the results. I find that masking the Galactic plane by 5 or 10 
degrees changes $\baobs$ in Fig. \ref{fig:sp_NT} by a tiny value $<0.008$. This is so tiny that the fit in 
Fig. \ref{fig:sp_S} remains unchanged.

A small fraction of radio source population is known to exhibit peaked-spectrum or convex spectrum between the 
TGSS and NVSS frequencies \citep{Callingham:2017}. These sources can be identified if the flux density is known 
in a large fractional bandwidth e.g. the GaLactic and Extragalactic All-sky Murchison Widefield Array (GLEAM) 
catalogue \citep{Hurley:2017gleam}. Ideally, I should mask all peaked-spectrum sources 
as these source do not  follow equation \ref{eq:aobs}. However, as we have flux densities only at two frequencies we can not 
identify peaked-spectrum sources. Nevertheless, I use \cite{Callingham:2017} peaked-spectrum source catalogue prepared from 
GLEAM survey, which covers $\sim60\%$ of TGSS-NVSS surveys, and identify in total 1069 peaked and 83 convex spectrum sources in 
our common catalogue. I mask these sources and find that the mean spectral index in Fig. \ref{fig:sp_NT} changes by $<0.003$.  
The remaining 40\% of TGSS-NVSS sky, not covered by GLEAM, is expected to have similar fraction of peaked-spectrum 
sources, i.e. around one thousand more peaked-spectrum sources, even so we do not expect any significant change in the results.
Furthermore we expect a few percent of mJy radio sources to be variable on the time-scale of years 
\citep{Oort:1985,Fan:2007}. The unresolved compact sources e.g. blazars show more than 4$\sigma$ flux 
varibility and constitutes a small faction of total radio population. For example 0.1\% of the unresolved sources from 
FIRST and NVSS  are found to be variable \cite{Ofek:2011}. Similarly \cite{Bell:2019} report that 
$\sim0.15$\% of bright compact sources at low frequencies ($>4$ Jy at 154 MHz) show significant long-term variability.
In our case variable sources may constitute not more than a few percentage of total population and may show very steep or 
flat spectral index and contribute to the outliers of histograms in Fig. \ref{fig:sp_NT} and \ref{fig:sp_NT2}, 
however, this is not expected to change the mean spectral index significantly.

\subsection{Radio spectral index dependence on size}
\label{ssc:res:Sp_Size} 

\begin{figure}[H]
\includegraphics[width=\wx\textwidth]{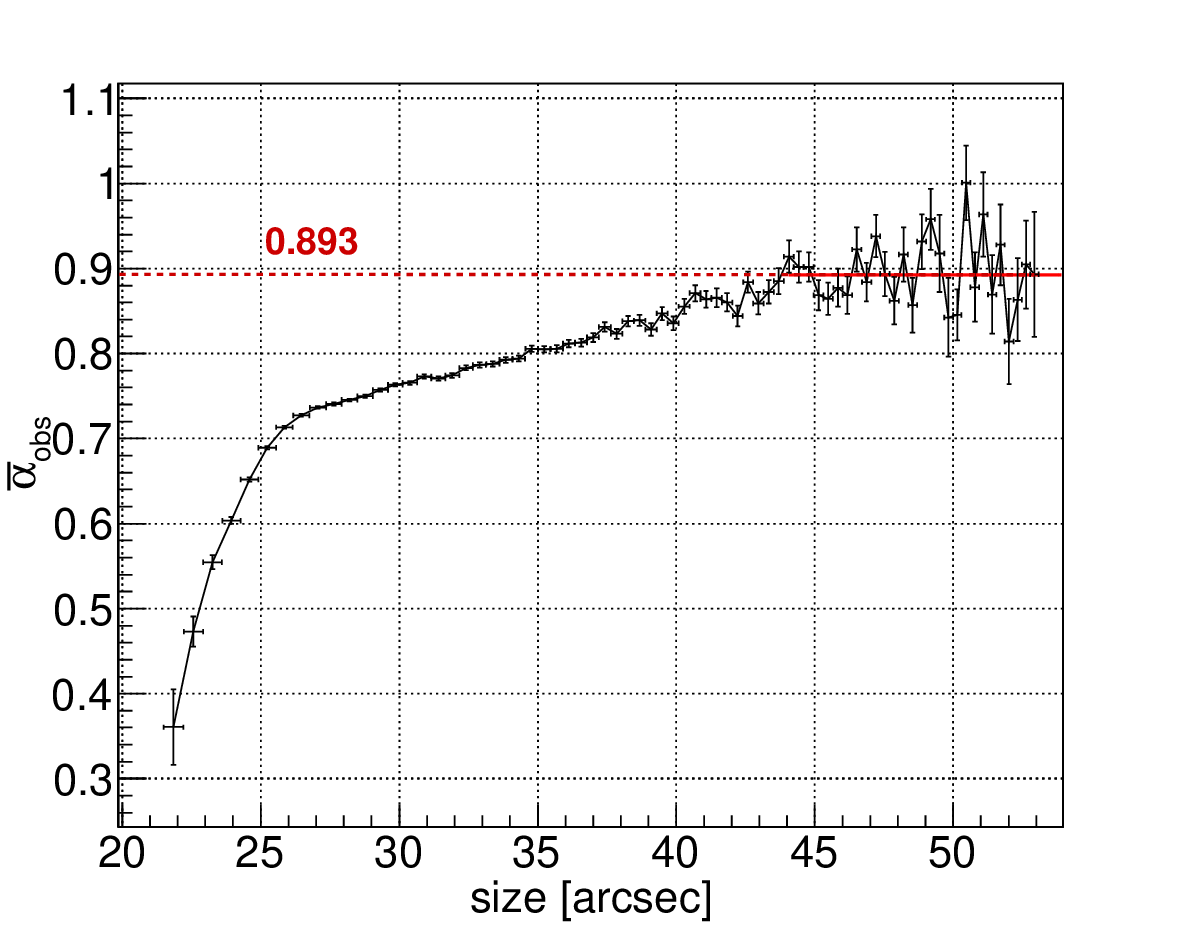}
\caption{ The mean radio spectral index, $\baobs$, dependence on size ($\sqrt{ab}$) for sources in the LF sample. 
The vertical and horizontal  error-bars represent the uncertainty in $\baobs$ and
the width of the  bin. The $\baobs$ dependence on size saturates very clearly for the sources above 
44 arcsec (roughly 340 kpc for  $z\approx0.8$). The red dashed line shows the mean $\aobs = 0.893\pm0.223$ 
from sources above 44 arcsec. }
\label{fig:aobs_LF}
\end{figure}

The size is defined as  $\sqrt{ab}$ (in arcsec) where $a$ and $b$ are, respectively, the major and minor axis of 
the Gaussian fit to the source  as given in TGSS ADR1 . As mentioned in Section \ref{sc:alp_obs}, the observed 
spectral index consists of radio emission from jets and lobes.  Fig. \ref{fig:aobs_LF} shows
$\aobs$ as a function of  size for LF sample.  The mean $\aobs$ is computed for sources lying in logarithmic 
bins of $\sqrt{ab}$.  The  spectral index increases rapidly with size until $\le26$\arcsec but flattens at 
larger sizes, becomes   nearly  independent of size beyond $\sim44$\arcsec and saturates to the value $\aobs=0.893\pm0.223$. 

\begin{figure}
\includegraphics[width=\wx\textwidth]{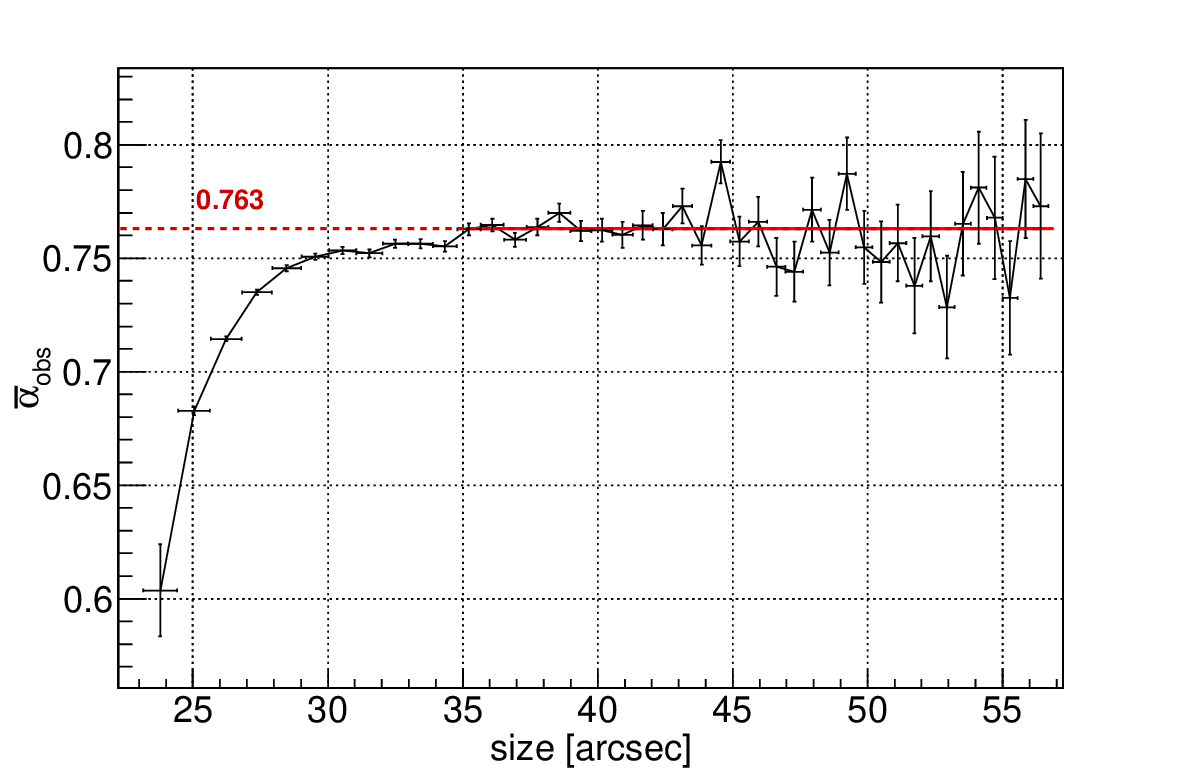}
\caption{ The same as the previous figure but for the HF sources. The red dashed line shows the mean
$\aobs$ from sources above 35 arcsec. The $\aobs$ above 35 arcesec for HF sample is $0.763\pm0.211$. }
\label{fig:aobs_HF}
\end{figure}

\begin{figure}
\includegraphics[width=\wx\textwidth]{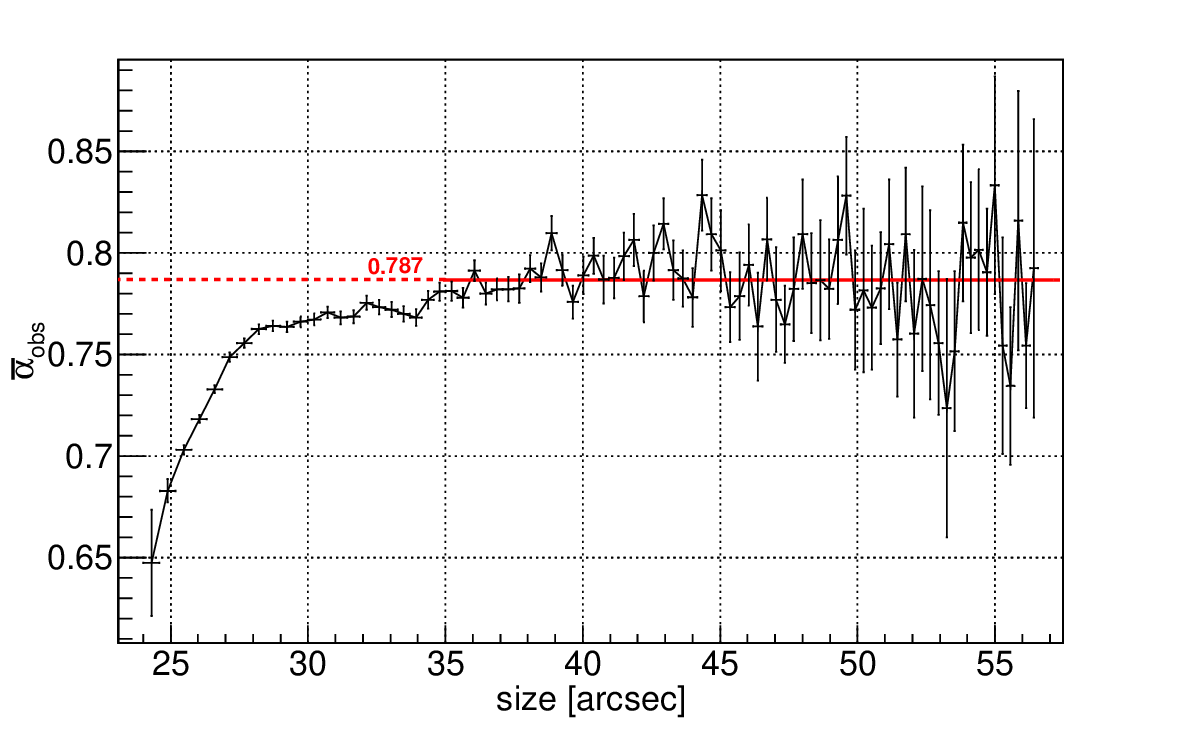}
\caption{ The same as the Fig. \ref{fig:aobs_LF} but for sources with TGSS flux density $>200$mJy.The red dashed line shows the mean
$\aobs$ from sources above 35 arcsec. The $\aobs$ above 35 arcesec is $0.787\pm0.205$.}
\label{fig:aobs_HF2}
\end{figure}

Fig. \ref{fig:aobs_HF} shows $\aobs$ versus $\sqrt{ab}$ for  the HF sample. The spectral index increases very 
rapidly with source size and saturates to the limiting value  $\aobs=0.763\pm0.211$ at large sizes ($\gtsim35$\arcsec). 
For ultra large sources, the emission is lobe dominated and therefore $\aobs$ represents the spectral index from lobes. Finally, I plot the $\aobs$ 
size dependence for sources above and equal ${\rm S_{\rm TGSS}} = 200$ mJy and ${\rm S_{NVSS}} = 40$ mJy in Fig. \ref{fig:aobs_HF2}. 
The spectral index  saturates to the limiting value  $\aobs=0.763\pm0.211$ above sizes $\gtsim35$\arcsec. For these high flux density 
limits we can be very certain about TGSS completness.

\section{Summary and Discussion}
\label{sc:summary}

I have computed the radio spectral index from a catalogue of  about half a million  common sources in 
NVSS and TGSS. The flux density accuracy in the common catalogue is better than 10\% which is sufficient 
for the studies performed in this paper. I check the robustness of the results with the Galactic plane cuts and find 
no significant change. I also mask the known peaked-spectrum sources identified from GLEAM survey covering $\sim 60\%$ 
of the common catalogue and find no significant change in the results.  We have relatively small number of sources 
close to survey resolution i.e. with size $<25$ arcsec ( $\sim$ 5\% and 1\%  of LF and HF sample, respectively) and 
some fraction of these small size sources may belong to Compact Steep Spectrum (CSS) and Gigahertz Peaked 
Spectrum  (GPS) population \citep{Odea:1998,Snellen:2000}. In particular, we observe the 
very flat spectral index ($<0.5$) for the very compact sources in LF sample, this is likely to be the result of 
significant CSS or GPS population \citep{Mahony:2016,Mahony:2016b}. Also, with extreme 
sizes (small or very large) the orientation is less likely to be randomly oriented. This all together may introduce 
some bias in the observed spectral index and the results may be relatively less reliable for sources with  extreme sizes.  
The saturated spectral index for both LF and HF sample is robust and independent of these possible systematics 
and biases though.

I divide the catalogue in two 
flux density limited samples LF and HF. 
The large number of sources enables us to study the spectral index in differential size bins. 
The spatially integrated radio spectral index, relevant to this work, consists of radio emission 
from lobes and jets. I explore the source size  dependence of observed spectral index and draw the 
following conclusions:
\begin{itemize}
\item I confirm the steepening of the radio spectral index in the low frequency bins
with increasing flux density 
\citep{Gopal-Krishna:1982,Steppe:1984,Windhorst:1993,Prandoni:2006,Ibar:2009,Ishwara:2010,Randall:2012}. 
This trend is inverted if we sort the sources in high frequency flux density bins.

\item  The spectral index becomes nearly independent of source size above $\sim44$ arcsec for LF sample and above $\sim35$ arcsec 
for HF sample.  The saturated spectral index for both LF and HF sample is in between $0.75-0.90$, consistent with previous 
studies \citep{Intema:2011,Williams:2013,Williams:2016,Hardcastle:2016,Mahony:2016b}. 
Both samples are hosts of AGNs, the LF sample is presumably dominated by FR I galaxies \citep{Wilman:2008}. 
Assuming that the sources lie at redshifts  $z\approx 0.8$, 44 arcsec size corresponds to $\sim$340 kpc 
(assuming standard $\Lambda$CDM cosmology and using cosmological parameters from \cite{Planck:CosmoPar}). 
For very compact sources in LF sample, the observed spectral index is flatter $\sim 0.4$, which hints 
of a significant thermal component from the nucleus in total radio emission for these compact sources. 
Alternatively, a significant fraction of these compact sources may belong to CSS and GPS population. 
The CSS and GPS sources may exhibiting a very flat 150 MHz - 1.4 GHz spectral index \citep{Mahony:2016,Mahony:2016b}. 
The LF sample is  increasingly incomplete for low flux densities, even so the results 
qualitatively represent the low flux density radio sources. 

\item HF sample exhibits  a very strong  dependence of the spectral index on size below $\sim35$ arcsec. This is likely 
due to an increasing fraction of jets and  nuclear radio emission in comparison to emission from lobes. The sources 
with larger sizes reach  a saturated  spectral index $\aobs=0.763\pm0.211$. The radio emission from these large sources 
is lobe dominated and so the observed spectral index can be seen as  characteristic of radio emission from lobes.  

\item The saturated  spectral index for LF sample is steeper than HF sample. The LF sample is 
predominantly FR I \citep{Wilman:2008} and the lobes in these galaxies exhibit steepest spectra \citep{Kembhavi:1999}, 
the electrons in lobes are most retarded and aged. In comparison with HF sample, we see a relatively slow rise of spectral 
index with size for LF sample. However, I remind the reader that the LF sample is increasingly incomplete for 
low flux densities and this slow rise of spectral index may be the result of incompleteness. The sharp saturation of 
spectral index for HF sample, which contains almost equal number of FR I and FR II galaxies \citep{Wilman:2008}, 
indicates very clear domination of radio emission from lobes for these bright galaxies. 

\end{itemize}

\section*{Acknowledgments}

I thank Adi Nusser and Ari Laor for discussion and their extensive and thoughtful comments on this work, and the 
anonymous referee for useful comments that helped to improve the paper. I especially   thank Huib T. Intema 
for help with the TGSS ADR1 catalogue. This work is  supported by NAOC youth talent fund 110000JJ01. 
I have used CERN ROOT 5.34/21 \citep{root} for generating plots. 
\bibliographystyle{raa}
\bibliography{master}

\begin{thebibliography}{54}
\providecommand\natexlab[1]{#1}
\providecommand\JournalTitle[1]{#1}

\bibitem[{Antonucci}(1993)]{Antonucci:1993}
{Antonucci}, R. 1993, \araa, 31, 473

\bibitem[{Bell}(1978)]{Bell:1978}
{Bell}, A.~R. 1978, \mnras, 182, 147

\bibitem[{Bell} {et~al.}(2019)]{Bell:2019}
{Bell}, M.~E., {Murphy}, T., {Hancock}, P.~J., {et~al.} 2019, \mnras, 482, 2484

\bibitem[{Biermann}(1976)]{Biermann:1976}
{Biermann}, P. 1976, \aap, 53, 295

\bibitem[{Blandford} \& {K{\"o}nigl}(1979)]{Blandford:1979}
{Blandford}, R.~D., \& {K{\"o}nigl}, A. 1979, \apj, 232, 34

\bibitem[{Blandford} \& {Ostriker}(1978)]{Blandford:1978}
{Blandford}, R.~D., \& {Ostriker}, J.~P. 1978, \apjl, 221, L29

\bibitem[{Bridle} \& {Perley}(1984)]{Bridle:1984}
{Bridle}, A.~H., \& {Perley}, R.~A. 1984, \araa, 22, 319

\bibitem[Brun {et~al.}(2001)]{root}
Brun, R., Rademakers, F., {et~al.} 2001, {ROOT web page,
  \texttt{http://root.cern.ch/}}

\bibitem[{Callingham} {et~al.}(2017)]{Callingham:2017}
{Callingham}, J.~R., {Ekers}, R.~D., {Gaensler}, B.~M., {et~al.} 2017, \apj,
  836, 174

\bibitem[{Condon}(1992)]{Condon:1992}
{Condon}, J.~J. 1992, \araa, 30, 575

\bibitem[Condon {et~al.}(1998)]{Condon:1998}
Condon, J.~J., Cotton, W.~D., Greisen, E.~W., {et~al.} 1998, AJ, 115, 1693

\bibitem[{Condon} \& {Yin}(1990)]{Condon:1990}
{Condon}, J.~J., \& {Yin}, Q.~F. 1990, \apj, 357, 97

\bibitem[{De Young}(1976)]{Young:1976}
{De Young}, D.~S. 1976, \araa, 14, 447

\bibitem[{Fan} {et~al.}(2010)]{Fan:2010}
{Fan}, J.-H., {Yang}, J.-H., {Tao}, J., {Huang}, Y., \& {Liu}, Y. 2010, \pasj,
  62, 211

\bibitem[{Fan} {et~al.}(2007)]{Fan:2007}
{Fan}, J.~H., {Liu}, Y., {Yuan}, Y.~H., {et~al.} 2007, \aap, 462, 547

\bibitem[{Fanaroff} \& {Riley}(1974)]{Fanaroff:1974}
{Fanaroff}, B.~L., \& {Riley}, J.~M. 1974, \mnras, 167, 31P

\bibitem[{Gopal-Krishna} \& {Steppe}(1982)]{Gopal-Krishna:1982}
{Gopal-Krishna}, \& {Steppe}, H. 1982, \aap, 113, 150

\bibitem[{Hardcastle} {et~al.}(2016)]{Hardcastle:2016}
{Hardcastle}, M.~J., {G{\"u}rkan}, G., {van Weeren}, R.~J., {et~al.} 2016,
  \mnras, 462, 1910

\bibitem[{Harwood} {et~al.}(2015)]{Harwood:2015}
{Harwood}, J.~J., {Hardcastle}, M.~J., \& {Croston}, J.~H. 2015, \mnras, 454,
  3403

\bibitem[{Harwood} {et~al.}(2013)]{Harwood:2013}
{Harwood}, J.~J., {Hardcastle}, M.~J., {Croston}, J.~H., \& {Goodger}, J.~L.
  2013, \mnras, 435, 3353

\bibitem[{Harwood} {et~al.}(2016)]{Harwood:2016}
{Harwood}, J.~J., {Croston}, J.~H., {Intema}, H.~T., {et~al.} 2016, \mnras,
  458, 4443

\bibitem[{Harwood} {et~al.}(2017b)]{Harwood:2017b}
{Harwood}, J.~J., {Hardcastle}, M.~J., {Morganti}, R., {et~al.} 2017b, \mnras,
  469, 639

\bibitem[{Hurley-Walker} {et~al.}(2017)]{Hurley:2017gleam}
{Hurley-Walker}, N., {Callingham}, J.~R., {Hancock}, P.~J., {et~al.} 2017,
  \mnras, 464, 1146

\bibitem[{Ibar} {et~al.}(2009)]{Ibar:2009}
{Ibar}, E., {Ivison}, R.~J., {Biggs}, A.~D., {et~al.} 2009, \mnras, 397, 281

\bibitem[{Intema} {et~al.}(2017)]{Intema:2016tgss}
{Intema}, H.~T., {Jagannathan}, P., {Mooley}, K.~P., \& {Frail}, D.~A. 2017,
  \aap, 598, A78

\bibitem[{Intema} {et~al.}(2011)]{Intema:2011}
{Intema}, H.~T., {van Weeren}, R.~J., {R{\"o}ttgering}, H.~J.~A., \& {Lal},
  D.~V. 2011, \aap, 535, A38

\bibitem[{Ishwara-Chandra} {et~al.}(2010)]{Ishwara:2010}
{Ishwara-Chandra}, C.~H., {Sirothia}, S.~K., {Wadadekar}, Y., {Pal}, S., \&
  {Windhorst}, R. 2010, \mnras, 405, 436

\bibitem[{Kellermann}(1966)]{Kellermann:1966}
{Kellermann}, K.~I. 1966, \apj, 146, 621

\bibitem[Kembhavi \& Narlikar(1999)]{Kembhavi:1999}
Kembhavi, A.~K., \& Narlikar, J.~V. 1999, Quasars and Active Galactic Nuclei An
  Introduction (Cambridge University Press)

\bibitem[{Kirk} {et~al.}(2000)]{Kirk:2000}
{Kirk}, J.~G., {Guthmann}, A.~W., {Gallant}, Y.~A., \& {Achterberg}, A. 2000,
  \apj, 542, 235

\bibitem[{Laing} \& {Bridle}(2013)]{Laing:2013}
{Laing}, R.~A., \& {Bridle}, A.~H. 2013, \mnras, 432, 1114

\bibitem[{Laing} \& {Bridle}(2014)]{Laing:2014}
{Laing}, R.~A., \& {Bridle}, A.~H. 2014, \mnras, 437, 3405

\bibitem[{Lal}(2013)]{GMRT}
{Lal}, D.~V. 2013, GMRT Observer's Manual, Available at
  \url{http://gmrt.ncra.tifr.res.in/gmrt_hpage/Users/doc/manual/Manual_2013/manual_20Sep2013.pdf}
  (November 18, 2013)

\bibitem[{Lemoine} \& {Pelletier}(2003)]{Lemoine:2003}
{Lemoine}, M., \& {Pelletier}, G. 2003, \apjl, 589, L73

\bibitem[{Mahony} {et~al.}(2016)]{Mahony:2016}
{Mahony}, E.~K., {Morganti}, R., {Prandoni}, I., {van Bemmel}, I., \& {LOFAR
  Surveys Key Science Project}. 2016, Astronomische Nachrichten, 337, 135

\bibitem[{Mahony} {et~al.}(2016b)]{Mahony:2016b}
{Mahony}, E.~K., {Morganti}, R., {Prandoni}, I., {et~al.} 2016b, \mnras, 463,
  2997

\bibitem[Nusser \& Tiwari(2015)]{Adi:2015nb}
Nusser, A., \& Tiwari, P. 2015, {ApJ}, 812, 85

\bibitem[{O'Dea}(1998)]{Odea:1998}
{O'Dea}, C.~P. 1998, \pasp, 110, 493

\bibitem[{Ofek} \& {Frail}(2011)]{Ofek:2011}
{Ofek}, E.~O., \& {Frail}, D.~A. 2011, \apj, 737, 45

\bibitem[{Oort} \& {Windhorst}(1985)]{Oort:1985}
{Oort}, M.~J.~A., \& {Windhorst}, R.~A. 1985, \aap, 145, 405

\bibitem[{Peterson}(1997)]{Peterson:1997}
{Peterson}, B.~M. 1997, {An Introduction to Active Galactic Nuclei} (Cambridge,
  New York Cambridge University Press)

\bibitem[{Planck Collaboration} {et~al.}(2016)]{Planck:CosmoPar}
{Planck Collaboration}, {Ade}, P.~A.~R., {Aghanim}, N., {et~al.} 2016, \aap,
  594, A13

\bibitem[{Prandoni} {et~al.}(2006)]{Prandoni:2006}
{Prandoni}, I., {Parma}, P., {Wieringa}, M.~H., {et~al.} 2006, \aap, 457, 517

\bibitem[{Randall} {et~al.}(2012)]{Randall:2012}
{Randall}, K.~E., {Hopkins}, A.~M., {Norris}, R.~P., {et~al.} 2012, \mnras,
  421, 1644

\bibitem[{Snellen} {et~al.}(2000)]{Snellen:2000}
{Snellen}, I.~A.~G., {Schilizzi}, R.~T., {Miley}, G.~K., {et~al.} 2000, \mnras,
  319, 445

\bibitem[{Steppe} \& {Gopal-Krishna}(1984)]{Steppe:1984}
{Steppe}, H., \& {Gopal-Krishna}. 1984, \aap, 135, 39

\bibitem[{Swarup}(1991)]{Swarup:1991}
{Swarup}, G. 1991, in Astronomical Society of the Pacific Conference Series,
  Vol.~19, IAU Colloq. 131: Radio Interferometry. Theory, Techniques, and
  Applications, ed. T.~J. {Cornwell} \& R.~A. {Perley}, 376

\bibitem[Tiwari \& Nusser(2016)]{Tiwari:2016adi}
Tiwari, P., \& Nusser, A. 2016, Journal of Cosmology and Astroparticle Physics,
  2016, 062

\bibitem[{Urry} \& {Padovani}(1995)]{Urry:1995}
{Urry}, C.~M., \& {Padovani}, P. 1995, \pasp, 107, 803

\bibitem[{White} {et~al.}(2005)]{White:2005}
{White}, R.~L., {Becker}, R.~H., \& {Helfand}, D.~J. 2005, \aj, 130, 586

\bibitem[{Williams} {et~al.}(2013)]{Williams:2013}
{Williams}, W.~L., {Intema}, H.~T., \& {R{\"o}ttgering}, H.~J.~A. 2013, \aap,
  549, A55

\bibitem[{Williams} {et~al.}(2016)]{Williams:2016}
{Williams}, W.~L., {van Weeren}, R.~J., {R{\"o}ttgering}, H.~J.~A., {et~al.}
  2016, \mnras, 460, 2385

\bibitem[{Wilman} {et~al.}(2008)]{Wilman:2008}
{Wilman}, R.~J., {Miller}, L., {Jarvis}, M.~J., {et~al.} 2008, \mnras, 388,
  1335

\bibitem[{Windhorst} {et~al.}(1993)]{Windhorst:1993}
{Windhorst}, R.~A., {Fomalont}, E.~B., {Partridge}, R.~B., \& {Lowenthal},
  J.~D. 1993, \apj, 405, 498

\end{thebibliography}
\end{document}